\documentclass[prb,twocolumn,amsfonts,amssymb,amsmath]{revtex4}

\usepackage[pdftex]{graphicx}
\usepackage{dcolumn}
\usepackage{bm}

\begin{document}

\title{Collective orbital excitations in orbitally ordered YVO$_3$ and HoVO$_3$}

\author{E.~Benckiser$^1$, R.~R\"{u}ckamp$^2$, T.~M\"{o}ller$^1$, T.~Taetz$^{3}$, A.~M\"{o}ller$^{3}$, A.~A.~Nugroho$^{4,5}$,
T.~T.~M.~Palstra$^5$, G.S. Uhrig$^6$, and M.~Gr\"{u}ninger$^{1}$}
\affiliation{$^1$II. Physikalisches Institut, Universit\"{a}t zu K\"{o}ln, Z\"{u}lpicher Str.\ 77, 50937 K\"{o}ln, Germany, \\
$^2$2.\ Physikalisches Institut A, RWTH Aachen, 52056 Aachen, Germany, \\
$^3$Institut f\"{u}r Anorganische Chemie, Universit\"{a}t zu K\"{o}ln, 50939 K\"{o}ln, Germany, \\
$^4$Jurusan Fisika, Institut Teknologi Bandung, Jl.\ Ganesha 10, Bandung 40132, Indonesia, \\
$^5$Zernike Institute for Advanced Materials, University of Groningen, Nijenborgh 4, 9747 AG Groningen, The Netherlands, \\
$^6$Theoretische Physik I, Technische Universit\"{a}t Dortmund, 44221 Dortmund, Germany}

\date{March 20, 2008}

\begin{abstract}
We study orbital excitations in the optical absorption spectra of YVO$_3$ and HoVO$_3$.
We focus on an orbital absorption band observed at 0.4\,eV for polarization $E\! \parallel \! c$.
This feature is only observed in the intermediate, monoclinic phase.
By comparison with the local crystal-field excitations in VOCl and with recent theoretical predictions
for the crystal-field levels we show that
this absorption band cannot be interpreted in terms of a local crystal-field excitation.
We discuss a microscopic model which attributes this absorption band to the exchange of two
orbitals on adjacent sites, i.e., to the direct excitation of two orbitons.
This model is strongly supported by the observed dependence on polarization and temperature.
Moreover, the calculated spectral weight is in good agreement with the experimental result.
\end{abstract}

\vskip2pc

\maketitle

\section{Introduction}

In strongly correlated transition-metal oxides, orbital interactions play a key role in
many intriguing phenomena such as the colossal magnetoresistance or the effective reduction
of dimensionality.\cite{tokura,khaliullinrev,khomskiirev,lee}
Orbitals on different sites interact with each other\cite{Kugel1973,Jahn1937} via the
collective Jahn-Teller effect, i.e., the coupling to the lattice, and via exchange interactions,
which are governed by the antisymmetrization of the total wave function including both the orbital
and the spin part. These interactions can result in coupled long-range spin and orbital order.
If the coupling to the lattice is dominant, the excitations are well described by
``local'' crystal-field (CF) excitations,\cite{Sugano,Ballhausen,Rueckamp2005} where ``local'' means
that the excitation can be treated as a change of the orbital occupation on a {\it single} site,
i.e., the dispersion is negligible.
In the opposite case of dominant exchange interactions, one expects novel collective elementary
excitations, namely orbital waves (orbitons) with a significant dispersion,\cite{Ishihara2000}
reflecting the propagation of the excited state. Thus orbitons are analogous to spin waves
-- propagating spin flips -- in a magnetically ordered state. Orbitons are expected to reveal
the fundamental orbital interactions responsible for the interesting physical properties.
In the quest for the experimental observation of orbitons, the central experimental task
is to demonstrate that the orbital exchange interactions are essential for the elementary
excitations. If this is the case, the excitations cannot be described in terms of single-site physics,
and we will use the term ``orbiton''.

The first claim for the observation of orbitons was based on Raman data of LaMnO$_3$,\cite{Saitoh}
but the relevant features later have been explained in terms of multi-phonons.\cite{Grueninger2002a}
In fact, in the manganites the orbital degree of freedom is connected with $e_g$ electrons, for
which the coupling to the lattice is strong in an octahedral environment. The vanadates \emph{R}VO$_3$
with two electrons occupying $t_{2g}$ orbitals may be considered as more promising
candidates.\cite{Ishihara2004,Khaliullin2001,deRay07} Recently, the observation of orbitons in Raman
data of \emph{R}VO$_3$ (R=Y, La, Nd) has been claimed at 43 and 62\,meV by
Miyasaka {\em et al.}\cite{Miyasaka2005,Miyasaka2006} and at 45 and 84\,meV by Sugai {\em et al.},\cite{Sugai2006}
but the proposed orbitons are hard to discriminate from (multi-)phonons and magnons,
and the assignment is controversial.\cite{Miyasaka2005,Miyasaka2006,Sugai2006}
Thus, an experimental proof for the existence of orbitons is still lacking.

\begin{figure}[b]
\center
\includegraphics[width=\linewidth, clip]{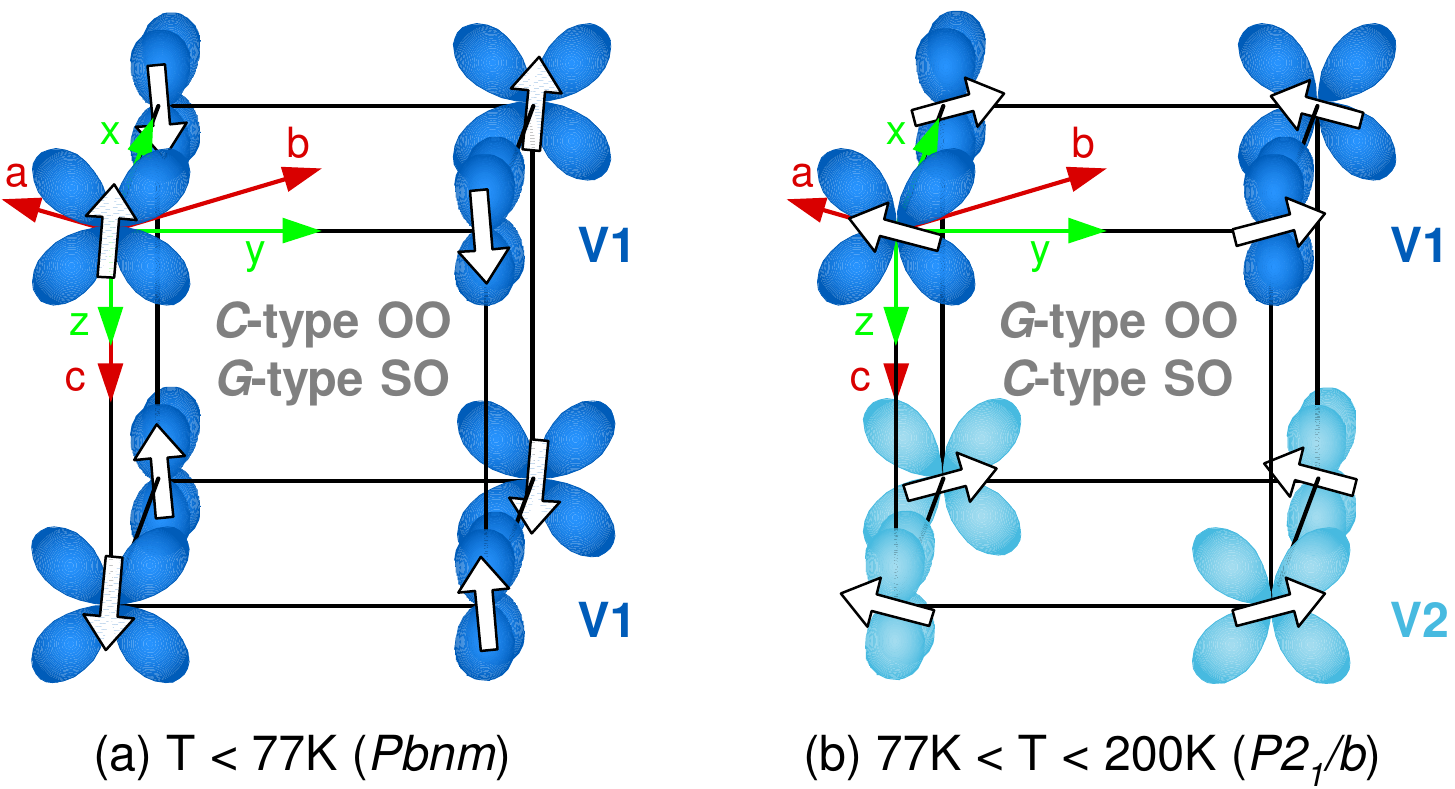}
\caption{Orbital and spin ordering patterns in YVO$_3$ for (a) $T_S\,<\,77$\,K and
(b) $T_S \,< \, T \, < \, T_{OO}$\,=\,
200\,K.\cite{Blake2002,Reehuis2006,Ren1998,Ren2000,Miyasaka2003,Tsvetkov2004,Noguchi2000,Sawada1998,Mizokawa99,Fang04,Otsuka06,Solovyev06,deRay07}
The $xy$ orbital is occupied by one electron on each site (not shown). The occupation of $xz$ and $yz$ orbitals
is \emph{C}-type below $T_S$ and \emph{G}-type above (see main text for more details).
Spin order (SO) is lost at $T_N$\,=\,116\,K.\@
The coordinates of the \textsl{Pbnm} and \textsl{P2$_1$/b} crystal systems are given in the upper left corner
of each figure.}
\label{Fig:YVO3orbitalOrdering}
\end{figure}

Here, we report on the observation of orbital excitations in the optical conductivity $\sigma(\omega)$
of orbitally ordered YVO$_3$ and HoVO$_3$. We focus on an absorption feature observed for $E\! \parallel \! c$
at about 0.4\,eV, well above the range of phonons and magnons and well below the Mott-Hubbard gap.
A comparison with the local CF excitations in the $3d^2$ system VOCl and with recent theoretical
results\cite{deRay07,Solovyev06} for the CF levels shows that this feature is hard to reconcile
with a local CF scenario, in particular as far as the energy, the polarization and temperature dependence
are concerned.
We discuss the microscopic exchange process and conclude that the feature at 0.4\,eV reflects the
exchange of two orbitals on neighboring sites, i.e., the direct excitation of two orbitons.

The vanadates \emph{R}VO$_3$ with \emph{R}\,=\,Y and Ho exhibit an orthorhombic crystal structure (\textsl{Pbnm})
at room temperature.\cite{Blake2002,Reehuis2006,Bombik1978} The undoped compounds represent Mott-Hubbard
insulators with two localized electrons in the $3d$ shell of each V$^{3+}$ ion. A crystal field of predominantly
octahedral symmetry
yields a splitting of the $3d$ states into a lower-lying, triply degenerate $t_{2g}$ level and a doubly degenerate
$e_g$ level. A detailed analysis of the structure reveals that the degeneracy of these levels is fully lifted
by an orthorhombic distortion of the VO$_6$ octahedra (\textsl{D}$_{2h}$ symmetry),\cite{Blake2002,Reehuis2006}
giving rise to a splitting of the $t_{2g}$ manifold into $xy$, $xz$, and $yz$ orbitals.
In YVO$_3$, a low-temperature orthorhombic phase (\textsl{Pbnm}) with \emph{G}-type spin order (SO) and
\emph{C}-type orbital order (OO) was found below
$T_S$\,=\,77\,K,\cite{Blake2002,Reehuis2006,Ren1998,Ren2000,Miyasaka2003,Tsvetkov2004,Noguchi2000,Sawada1998,Mizokawa99,Fang04,Otsuka06,Solovyev06,deRay07}
i.e., the $xy$ orbital is occupied at each V site, whereas the occupation of $xz$ and $yz$
orbitals alternates within the $ab$ plane (see Fig.\ \ref{Fig:YVO3orbitalOrdering}).
At $T_S$\,=\,77\,K a first-order structural phase transition to an intermediate, monoclinic phase with
\textsl{P2$_1$/b} symmetry has been observed. This monoclinic phase shows two different vanadium sites, V(1) and V(2),
which alternate along the $c$ axis. Therefore, the mirror symmetry perpendicular to the $ab$ plane is broken
in the intermediate phase. The spin ordering pattern changes from G-type below $T_S$ to C-type above.
Long-range magnetic order is lost at $T_N$\,=\,116\,K.\@
The structural phase transition from the monoclinic phase to the orthorhombic room-temperature phase is
observed at $T_{OO}$\,$\approx$\,200\,K, which has been interpreted as the long-range orbital ordering
temperature.\cite{Blake2002}
However, synchrotron x-ray diffraction data give evidence for the presence of OO up to about 300\,K.\cite{Noguchi2000}
For the intermediate monoclinic phase, it has been discussed controversially whether the physics has to be
described in terms of 'classical' orbital order or quantum orbital fluctuations:
It has been proposed that the intermediate phase of YVO$_3$ represents the first realization of a one-dimensional
orbital liquid and of an orbital Peierls phase with V(1)-V(2) orbital
dimers.\cite{Khaliullin2001,Ulrich2003,Horsch2003,Miyashita2004,Oles2007}
This has been challenged by LDA+$U$ and LDA+DMFT studies,\cite{Fang04,deRay07} which for YVO$_3$ find
orbital order and that at least below 300 K orbital quantum fluctuations are suppressed in YVO$_3$
by a sizeable ligand-field splitting.
The orbital ordering pattern in the monoclinic phase has been reported as G-type based on, e.g., resonant
x-ray diffraction,\cite{Noguchi2000} an analysis of the V-O bond lengths,\cite{Ren1998} or LDA+$U$ calculations.\cite{Fang04}
In comparison to C-type OO, the orbitals of every second layer along $c$ are shifted along $x$,
thus $xz$ and $yz$ alternate along $x$, $y$, and $z$ for G-tye OO (see Fig.\ \ref{Fig:YVO3orbitalOrdering}).
According to Hartree-Fock calculations, the size of GdFeO$_3$-type distortions is decisive for the
choice between G-type and C-type orbital order.\cite{Mizokawa99}
A recent LDA+DMFT study by De Raychaudhury {\it et al.} finds that the OO pattern is intermediate between C-type
and G-type due to the GdFeO$_3$ distortions, almost C-type in the intermediate phase.\cite{deRay07}
However, Noguchi {\it et al.}\cite{Noguchi2000} claim that their synchrotron x-ray diffraction data are fitted best by
G-type OO, but that the partial occupation of other orbitals is possible.
The compound HoVO$_3$ behaves very similar to YVO$_3$ with slightly different phase-transition temperatures
of $T_{OO}$\,$\approx$\,188\,K, $T_N$\,=\,114\,K, and $T_S$\,$\approx$\,40\,K.\cite{Bombik1978,Blake2008}
Detailed neutron and synchrotron x-ray scattering experiments were unable to establish the presence of
orbital fluctuations in HoVO$_3$.\cite{Blake2008}

\section{Experiment}

\begin{figure}[t]
\includegraphics[clip,width=0.95\linewidth]{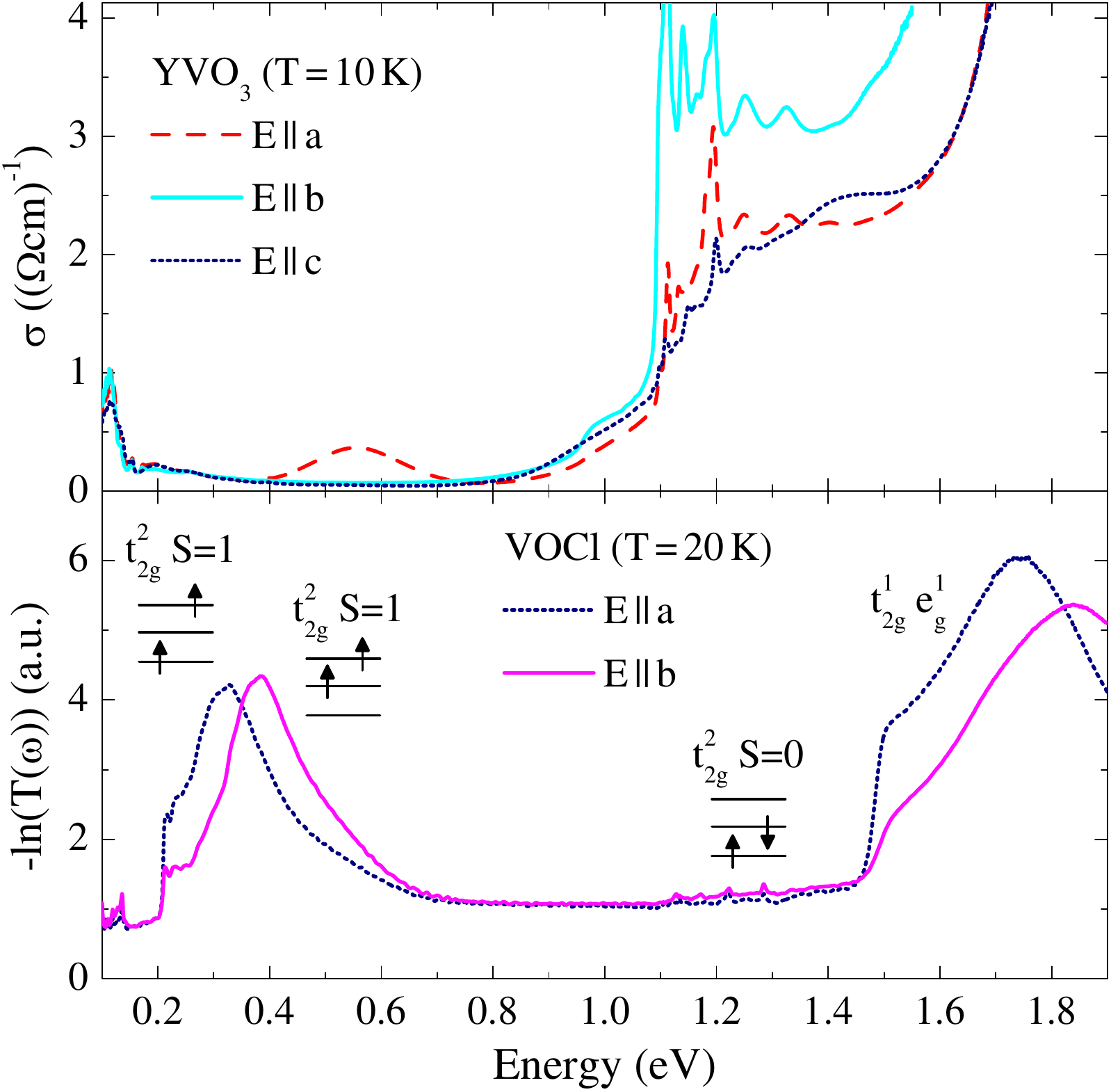}
\caption{(color online) Optical conductivity of YVO$_3$ in the low-temperature orthorhombic phase at $T$\,=\,10\,K
for $E$\,$\parallel$\,$a$, $b$, and $c$ (top panel) and
$-\ln({\rm T}(\omega))$ of VOCl at 20\,K for $E$\,$\parallel$\,$a$ and $b$ (in \textsl{Pmmn}; bottom
panel).
Three different orbital excitations within the $t_{2g}$ subshell are sketched in the lower panel. }
\label{fig:YVO3_VOCl}
\end{figure}

Single crystals of \emph{R}VO$_3$ with \emph{R}\,=\,Y and Ho have been grown by the
traveling-solvent floating-zone method.\cite{Blake2002}
The purity, stoichiometry and single-phase structure of the crystals was checked by x-ray diffraction
and thermogravimetry.
Typical crystal dimensions are a few mm along all three crystallographic axes.
The optical conductivity $\sigma(\omega)$ of YVO$_3$ was determined by measuring both the transmittance
and the reflectance\cite{Grueninger2002} between 0.06 and 1.9\,eV using a Fourier spectrometer.
The measurements have been performed using linearly polarized light with the electric field parallel
to the orthorhombic axes, i.e., $E$\,$\parallel$\,$a$, $b$, and $c$ (see top panel of Fig.~\ref{fig:YVO3_VOCl}).
For convenience, we use the same set of axes at all temperatures, i.e., we neglect the monoclinic
distortion of the structure.
This is justified because the monoclinic angle $\alpha$\,=\,89.98$^{\circ}$ is very close
to $90^\circ$.\cite{Blake2002}
The reflectance was measured on samples with a thickness of $d$\,$>$\,2\,mm in order to avoid backside
reflections. The transmittance data were collected
on a series of crystals with different thickness
(100\,$\mu$m\,$<$\,$d$\,$<$\,500\,$\mu$m), which were polished on both sides.

Single crystals of VOCl have been grown by the
chemical-vapor transport technique.
The purity of the crystals was checked by x-ray powder diffraction. Typical crystal dimensions are a
few mm$^2$ in the $ab$ plane and 10--100~$\mu$m along the $c$ axis.
In the case of VOCl and HoVO$_3$, we have measured the transmittance only.
The transmittance ${\rm T}(\omega)$ is a very sensitive probe for the determination of weakly infrared-active
excitations below the gap of these Mott-Hubbard insulators, where the reflectance is nearly
constant and featureless. Therefore, the absorption coefficient $\alpha(\omega) \propto -\ln({\rm T}(\omega))/d$
can be used equivalently to $\sigma(\omega)$ for the determination of weak orbital
excitations.\cite{Grueninger2002}

\section{Results and Discussion}

\begin{figure}[t]
\includegraphics[clip,width=0.99\linewidth]{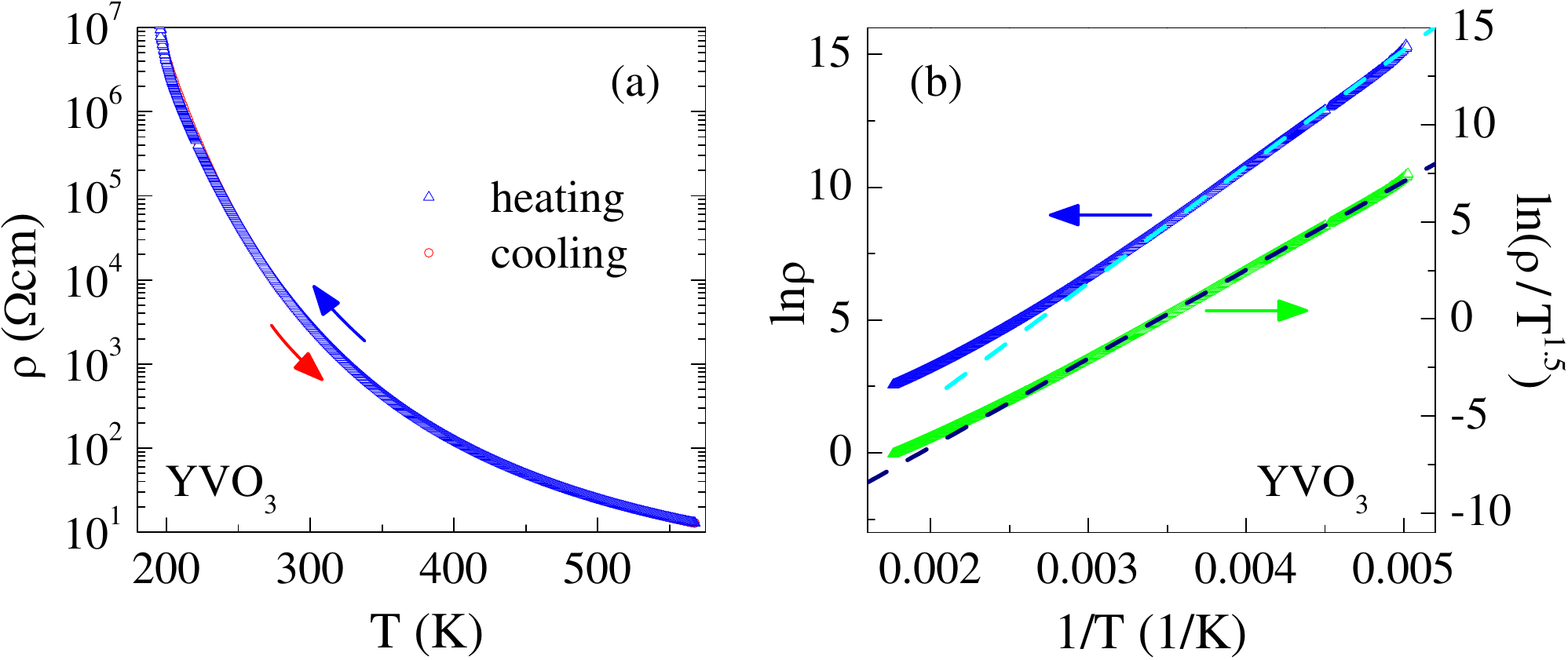}
\caption{(color online) (a) Resistivity $\rho$ of YVO$_3$ from $T$\,=\,200\,K up to 570\,K.\@
(b) $\ln{(\rho)}$ vs.\ 1/$T$ (left axis) and $\ln{(\rho / T^{\frac{3}{2}})}$ vs.\ $1/T$ (right axis).
Dashed lines show fits for activated behavior (left axis) and within the non-adiabatic
small-polaron model (right).
}
\label{fig:resistivity}
\end{figure}

The top panel of Fig.~\ref{fig:YVO3_VOCl} shows the optical conductivity $\sigma(\omega)$ of YVO$_3$ at 10\,K
in the transparent window of the Mott-Hubbard insulator, i.e., above the phonon range and below the electronic
interband excitations.
The lowest electronic transition corresponds to an excitation across the Mott-Hubbard gap, i.e., the transfer of
one electron from a $3d^2$ V$^{3+}$ site to another one, $3d^2\,3d^2 \rightarrow 3d^1\,3d^3$.
In the optical conductivity, the lowest electronic excitation is observed at 1.8\,eV [\onlinecite{Miyasaka2002,Tsvetkov2004}],
and a value of 1.6\,eV has been reported for the Mott-Hubbard gap based on a combination of ellipsometry and
LSDA+\textsl{U} calculations.\cite{Tsvetkov2004}
This agrees with the steep increase of $\sigma(\omega)$ in our data (see top panel of Fig.\ \ref{fig:YVO3_VOCl}).
However, the transmittance reveals that the very onset of excitations across the Mott-Hubbard gap is
somewhat lower. The precise onset is obscured by the superposition of spin-forbidden orbital excitations
between about 1.0 and 1.5\,eV (see below).
Charge-transfer excitations, involving the transfer of an electron
between V and O ions ($d^2$\,$\rightarrow$\,$d^3\bar{L}$),
are located above about 4\,eV.\cite{Tsvetkov2004}
Absorption features below the Mott-Hubbard gap have to be attributed to
phonons, magnons, excitons, orbital excitations, or to localized carriers trapped by impurities.
We exclude the latter for a number of reasons:
(i) the spectra of different samples of YVO$_3$ are identical,
(ii) the spectra of YVO$_3$ and HoVO$_3$ are very similar (see below),
(iii) the polarization and temperature dependence (see below), and (iv) the DC resistivity,
which is very large, 3$\cdot 10^6\, \Omega$cm at 200\,K.\@
In earlier experiments, an activation energy of $\Delta_{\rm act}$\,=\,0.25\,eV was obtained for polycrystalline
samples from resistivity data for 180\,K$<$\,$T$\,$<$\,300\,K,\cite{Kasuya1993} corresponding to an optical gap
of $2\Delta_{\rm act}$\,=\,0.5\,eV.\@
In Fig.\ \ref{fig:resistivity}(a) we plot $\rho(T)$ of our single crystals between 200 and 570\,K.
Note that $\rho(T)$ is about an order of magnitude larger than reported in Ref.\ \onlinecite{Kasuya1993}.
The Arrhenius plot in Fig.\ \ref{fig:resistivity}(b) suggests $2\Delta_{\rm act}\! \approx\! 0.8\,$eV,
but the data are not very well described by simple activated behavior.
The data are best described by the non-adiabatic small-polaron model,\cite{Holstein1959}
predicting $\rho$\,=\,$CT^{\frac{3}{2}}\exp{\left(\Delta_{\rm act} / (k_BT)\right)}$
(see \ Fig.~\ref{fig:resistivity}(b)). Within this model we obtain
$2\Delta_{\rm act}$\,=\,0.78\,eV, in agreement with both the estimate based on simple activated behavior
and in particular with the onset of absorption observed in $\sigma(\omega)$ (see top panel of
Fig.\ \ref{fig:YVO3_VOCl}). This value is significantly larger than reported in
Ref.\ \onlinecite{Kasuya1993}, demonstrating the high quality of our samples.
Remarkably, DFT-PIRG calculations\cite{Otsuka06} predict an indirect gap at 0.7\,eV, in excellent
agreement with our data. Altogether, we have strong evidence that all absorption features observed
in $\sigma(\omega )$ below 0.8\,eV arise from phonons, magnons, excitons, or orbital excitations.

The optical conductivity of YVO$_3$ below 0.9\,eV is given in Fig.\ \ref{fig:YVO3}.
The absorption due to phonons gives rise to the steep increase of $\sigma(\omega)$
below about 80\,meV, in agreement with the data reported in Ref.\ \onlinecite{Tsvetkov2004}.
Thus, weak two-phonon features can be observed up to about 160\,meV, whereas three-phonon
absorption is expected to be still much weaker.
A contribution to $\sigma(\omega)$ from spin waves may arise in the form of
two-magnon-plus-phonon absorption.\cite{Lorenzana1995} However, in YVO$_3$
the spin-wave dispersion does not exceed 40\,meV.\cite{Ulrich2003}
Thus, a possible two-magnon-plus-phonon contribution is expected to peak below 0.2\,eV and
clearly is not related to the features observed above 0.2\,eV.\@

As far as excitons are concerned, we have to distinguish between strongly and weakly bound excitons.
In a Mott-Hubbard insulator, an exciton is a bound state of an electron in the upper Hubbard band
(i.e., a double occupancy) and a hole in the lower Hubbard band (an ``empty site'').
In the case of a weakly bound exciton, the electron and the hole occupy distinct sites, e.g., nearest-neighbor
sites. The binding may result from the nearest-neighbor Coulomb attraction.
Recently, an excitonic resonance was reported\cite{goessling} in the orbitally
ordered $3d^1$ Mott-Hubbard insulator YTiO$_3$ at 1.95\,eV.\@ We consider it as unrealistic to
assume a binding energy large enough to pull such a weakly bound exciton far below 1\,eV in YVO$_3$.
Moreover, the spectral weight of the exciton in YTiO$_3$ is more than two orders of magnitude larger
than the spectral weight between 0.1 and 0.7\,eV in YVO$_3$.

\subsection*{Orbital excitations}
The physics is different for a strongly bound exciton, in which case the electron and the hole share
the $3d$ shell of the {\em same} site. This actually corresponds to an orbital or $d$-$d$ excitation
(or to a spin excitation, which has been ruled out above). These excitations may occur far below the gap
because the binding energy and the gap are both of the order of the on-site Coulomb repulsion $U$.
In other words, double occupancy is omitted if the electron and the hole share the same site,
thus one does not have to pay the energy $U$.
Orbital excitations in the form of local crystal-field excitations are a common
feature in many transition-metal compounds in the considered frequency
range.\cite{Sugano,Ballhausen,Rueckamp2005} The central question to us is whether the observed orbital
excitations are such local crystal-field excitations or propagating orbitons, reflecting the importance
of orbital exchange interactions. The orbiton dispersion for orbitally-ordered vanadates has been
investigated theoretically.\cite{Ishihara2004}
Predictions for Raman scattering\cite{Ishihara2004,Miyasaka2005,Miyasaka2006} and inelastic neutron
scattering\cite{Ishihara2004} have been discussed,
but contributions to $\sigma(\omega)$ for the case of dominant exchange interactions have not been
considered thus far. Moreover, a quantitative description requires that both the exchange interactions and the
coupling to the lattice are treated on the same footing,\cite{Schmidt2007,Brink2001}
but up to now such calculations have not been reported for the vanadates.
Therefore, we first compare our results with the expectations for local crystal-field
excitations of $3d^2$ V$^{3+}$ ions.
We use VOCl as a typical example for the absorption spectrum of V$^{3+}$ ions in a predominantly
octahedral, but distorted environment (see bottom panel of Fig.~\ref{fig:YVO3_VOCl}).
In the sister compound TiOCl, the orbital excitations are very well described in terms of local
crystal-field excitations.\cite{Rueckamp2005,Rueckamp2005a}

\begin{figure}[t]
\includegraphics[clip,width=0.95\linewidth]{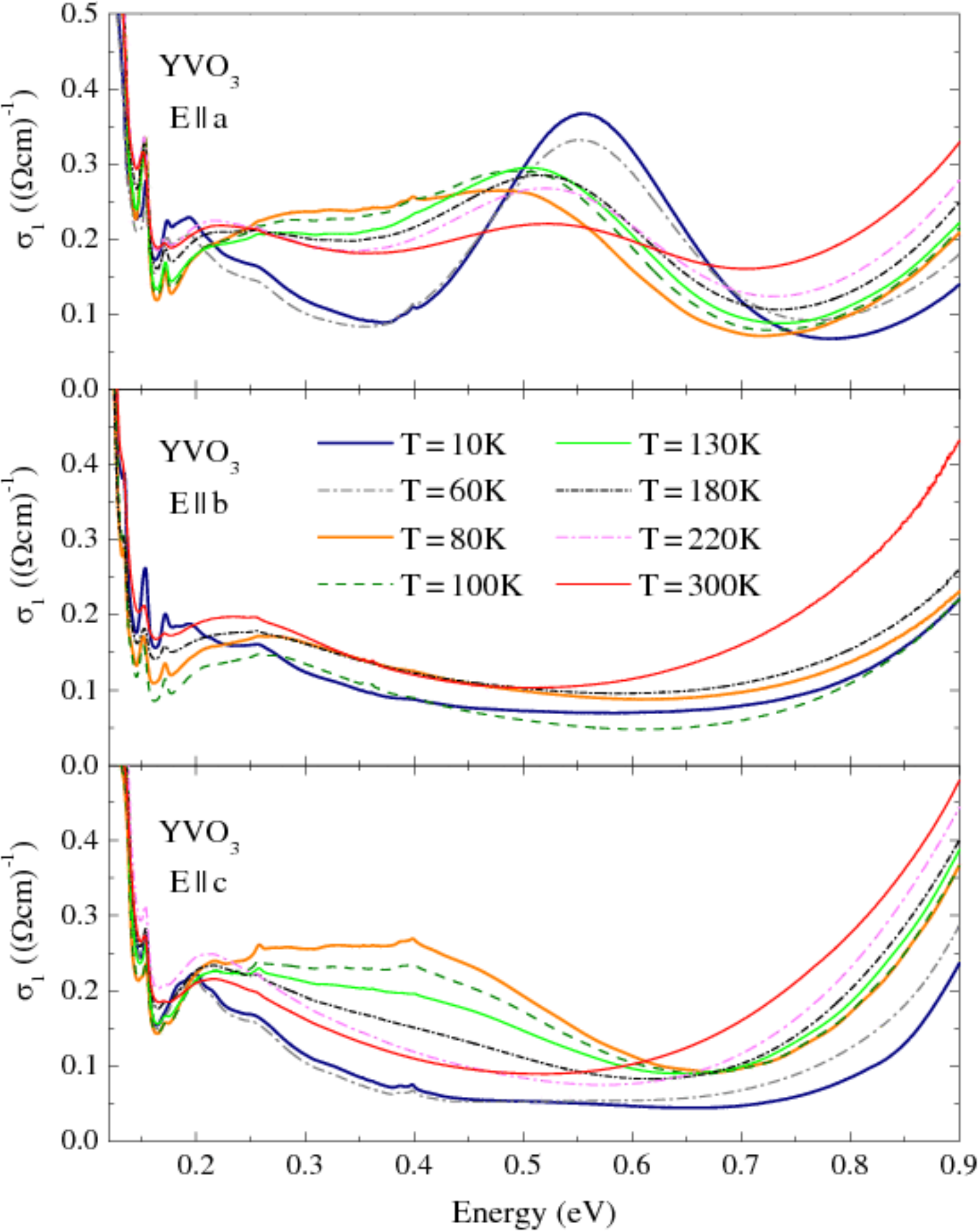}
\caption{(color online)
Temperature dependence of $\sigma(\omega)$ of YVO$_3$ in the mid-infrared range
for $E$\,$\parallel$\,$a$ (top panel), $E$\,$\parallel$\,$b$ (middle), and $E$\,$\parallel$\,$c$ (bottom).
The features at 0.4\,eV for $E$\,$\parallel$\,$c$ and 0.55\,eV for $E$\,$\parallel$\,$a$ constitute our
main experimental result.
} \label{fig:YVO3}
\end{figure}

YVO$_3$ shows inversion symmetry on the V sites, thus local CF excitations are not infrared-active
due to the parity selection rule. However, they become weakly allowed by the
simultaneous excitation of a symmetry-breaking phonon.\cite{Sugano,Ballhausen,Rueckamp2005}
In VOCl (\emph{Pmmn}) there is no inversion symmetry on the V site\cite{Ehrlich1959,VOCl} and the
CF excitations are weakly allowed without the additional excitation of a phonon.

For the sake of simplicity, we start from an undistorted octahedral crystal field,
the effect of a lower symmetry will be discussed below. The ground state of a $d^2$ system in
an octahedral field is the nine-fold degenerate $^3T_1$ level with total spin
$S$\,=\,1, in which two electrons with parallel spins occupy the $t_{2g}$ level, $t_{2g}^{\uparrow\uparrow}$.\cite{Sugano}
The splitting between $t_{2g}$ and $e_g$ levels typically amounts to $\gtrsim 2$\,eV for O$^{2-}$
ligands.\cite{Sugano,Ballhausen} For VOCl, this splitting is reduced due to the smaller
ligand strength of the Cl$^{1-}$ ions
and can be identified with the feature observed around 1.7\,eV (see bottom panel of Fig.\ \ref{fig:YVO3_VOCl}).
For YVO$_3$, it is reasonable to assume that the $t_{2g}$-$e_g$ splitting is larger than the Mott-Hubbard gap.
In the following, we focus on the excitations within the $t_{2g}$ shell, which are located at lower energies.

In an octahedral field, the spin-flip excitation from the $^3T_1$ ground state with $S$\,=\,1 to the
five-fold degenerate $S$\,=\,0 state ($^1T_2$, $^1E$; $t_{2g}^{\uparrow\downarrow}$) occurs at 2$J_H$.
The Hund exchange $J_H$\,$\approx$\,0.7\,eV (Ref.\ \onlinecite{Mizokawa96}) is hardly screened in a solid,
therefore this excitation is observed at very similar energies in different V
compounds,\cite{Bussiere01,Ishii02,Tregenna04} irrespective of the crystal structure.
Typical examples are the weak, sharp features observed between 1.1 and 1.3\,eV in VOCl (see bottom panel of
Fig.\ \ref{fig:YVO3_VOCl}). These spin-flip excitations are very weak due to the spin-selection rule,
they become weakly allowed by spin-orbit coupling or by the simultaneous excitation of a magnon.\cite{Sugano}
Such spin-flip bands often are very sharp because the orbital occupation stays the same, thus the coupling
to the lattice is only weak (the large width of spin-allowed excitations is attributed to vibronic
Franck-Condon sidebands). Also in YVO$_3$, sharp features are observed between 1.1 and 1.4\,eV which
can be attributed to spin-forbidden excitations. The oscillator strength is clearly enhanced compared
with VOCl. Presumably, this is due to the overlap with the onset of excitations across the Mott-Hubbard
gap. Mixing the two kinds of excitations will transfer some weight to the orbital bands, a process
called ``intensity stealing''.

In a local crystal-field scenario, all features observed significantly below 1\,eV have to be interpreted
as spin-conserving transitions within the $t_{2g}$ shell, i.e, both the ground state and the excited state
show $S$\,=\,1. Both in YVO$_3$ and in VOCl, the symmetry on the V sites is lower than tetragonal, thus
the $t_{2g}$ level is split into three distinct orbitals. For two electrons with parallel spins, there are
three distinct energy levels, each showing three-fold spin degeneracy. In strongly distorted VOCl we
observe the corresponding excitations at 0.3--0.4\,eV (see Fig.\ \ref{fig:YVO3_VOCl}).
For YVO$_3$, one expects lower excitation energies because the distortions away from an octahedral
environment are smaller than in VOCl.
Indeed, recent first-principles\cite{Solovyev06} and LDA+DMFT calculations\cite{deRay07} predict
intra-$t_{2g}$ excitations in YVO$_3$ in the range of 0.10 -- 0.20\,eV and 0.06 -- 0.24\,eV, respectively.
Our data of $\sigma(\omega)$ of YVO$_3$ show an absorption band centered around 0.20 -- 0.25\,eV at
all temperatures and for all polarization directions (see Fig.\ \ref{fig:YVO3}), in agreement
with these expectations.

\begin{figure}[b]
\includegraphics[clip,width=0.9\linewidth]{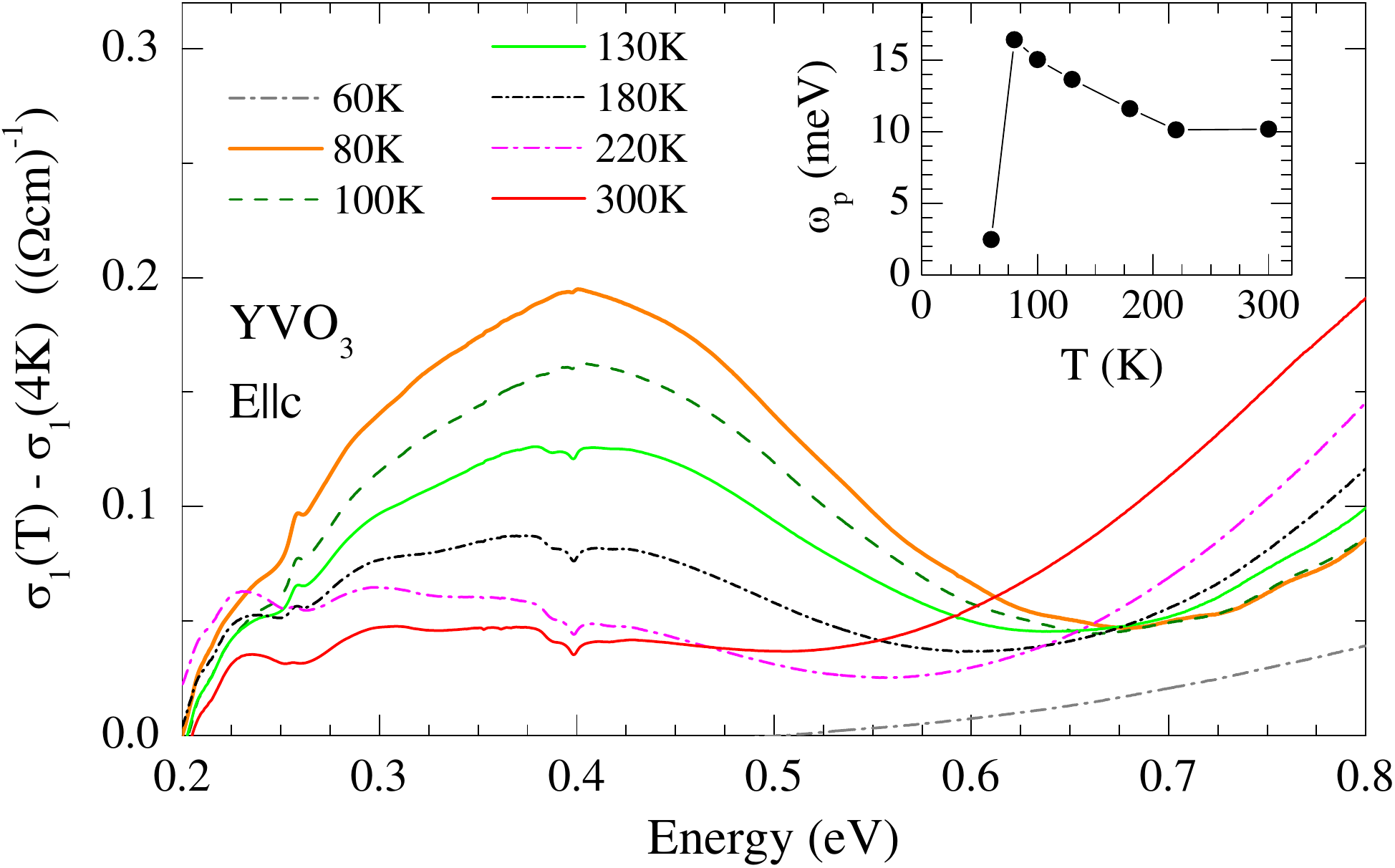}
\caption{(color online) Spectral weight and line shape of the peak at 0.4\,eV for $E$\,$\parallel$\,$c$ in YVO$_3$.
According to our microscopic model, the two-orbiton excitation is forbidden below $T_S = 77$\,K.\@
Therefore, the data for 4\,K have been subtracted for each temperature as an estimate for the background of, e.g.,
multi-phonon absorption. The inset depicts the plasma frequency $\omega_p$ (see Eq.\ \ref{eq:omegap})
for $\omega_{\rm lo}$\,=\,0.20\,eV and $\omega_{\rm hi}$\,=\,0.68\,eV.
} \label{fig:weightc}
\end{figure}

Our main experimental result is the observation of the two remaining features which {\em cannot} be explained within
a local crystal-field scenario, namely the peaks at about 0.4\,eV for $E$\,$\parallel$\,$c$ and at 0.55\,eV for
$E$\,$\parallel$\,$a$ (see Figs.\ \ref{fig:YVO3} and \ref{fig:weightc}). As follows from the discussion above,
these features are hard to reconcile with a local crystal-field scenario: the energy is too low for a spin-flip
transition and too high for a spin-conserving intra-$t_{2g}$ transition. Moreover, the strong polarization dependence
of both peaks is entirely unexpected for a phonon-assisted crystal-field excitation.
Since the phonon polarization is arbitrary, one does not expect strict polarization selection rules.\cite{Rueckamp2005}
Even in VOCl, where CF excitations are directly allowed, the absorption is very similar for the two polarization
directions. Finally, also the pronounced temperature dependence is unexpected for a local CF excitation.
For $E\! \parallel \! c$, the spectral weight around 0.4\,eV is independent of temperature for $T > T_{OO}$\,=\,200\,K,
increases with decreasing temperature below $T_{OO}$, and abruptly disappears at $T_S$\,=\,77\,K
(see inset of Fig.\ \ref{fig:weightc}).
For a local CF excitation, both the spectral weight and the energy in principle may change across a
structural phase transition.
However, LDA+DMFT calculations\cite{deRay07} predict that the intra-$t_{2g}$ excitation energies change by less
than 40 meV across $T_S$, thus the abrupt change of $\sigma(\omega)$ at $T_S$ cannot be explained.

\begin{figure}[t]
\includegraphics[clip,width=0.85\linewidth]{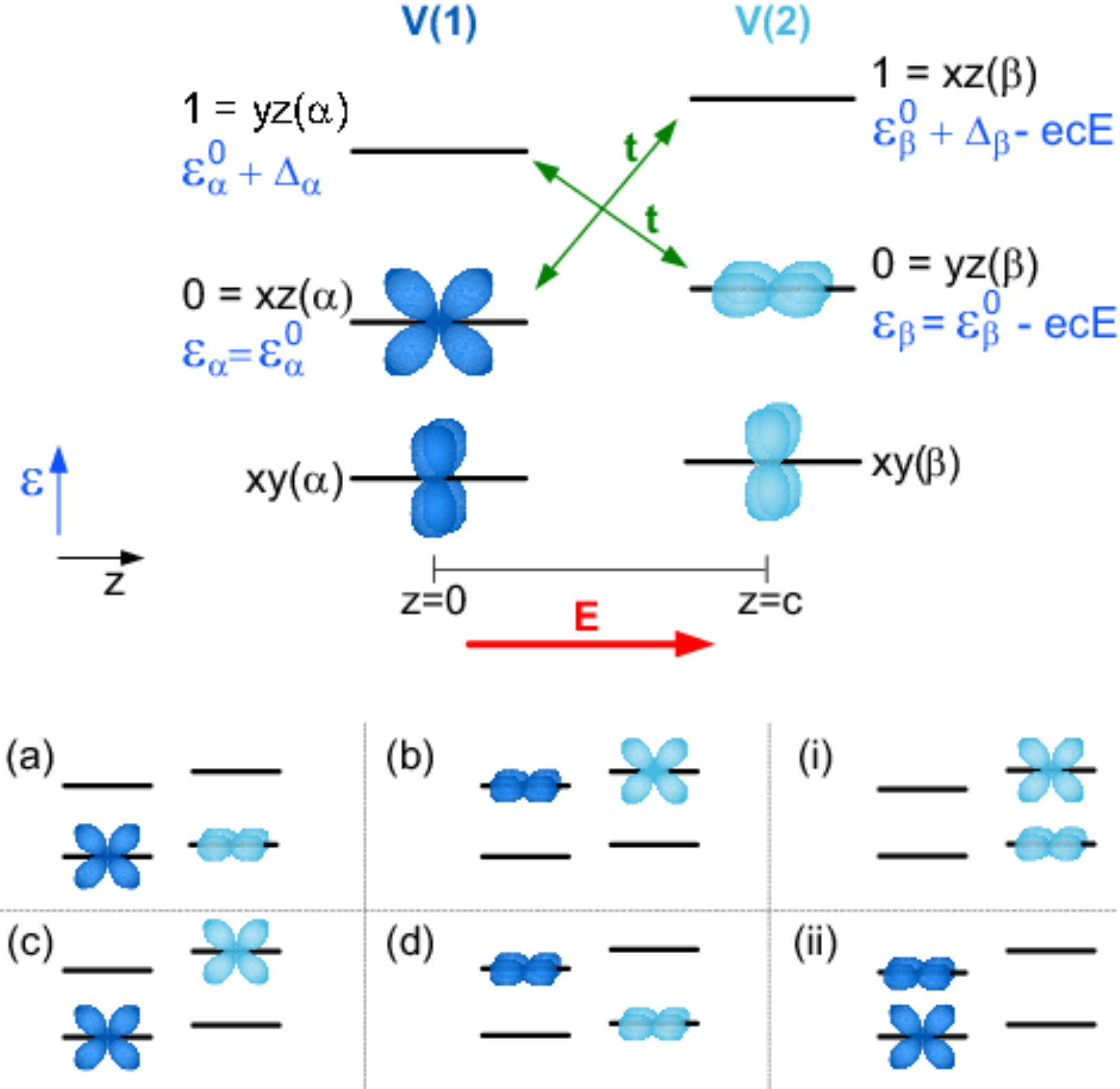}
\caption{(color online)
Top: Sketch of the local energy levels and of the hopping processes between neighboring V sites
for the two-orbiton excitation in the monoclinic phase of \emph{R}VO$_3$ with an electric field $E$ applied parallel
to the $c$ axis. Only $xz$ and $yz$ orbitals are considered for the exchange because the lower-lying $xy$ orbital
is always occupied by one electron. The spin is omitted since we consider fully polarized electrons.
See text for more details.
Bottom: For two electrons with parallel spins, there are six possibilities to occupy the $xz$ and $yz$ orbitals
on sites V(1) and V(2): (a) orbital ground state; (b) two-orbiton excited state; (c) \& (d) one-orbiton excited
states; (i), (ii) states in the upper Hubbard band with a doubly occupied V site.
} \label{fig:TwoOrbiton}
\end{figure}

\subsection*{Orbitons}
Thus far we have neglected the exchange interactions between orbitals on neighboring sites, which change
the character of the excitations from ``local'' CF excitations to propagating orbitons.
Here, we consider two different processes for the excitation of orbitons.\cite{Ishihara2004}
First, the excitation process itself may be based on the exchange of orbitals between adjacent sites (see below).
Second, an orbiton may be excited {\em locally} by flipping an orbital on a single site, e.g., from $xz$ to $yz$,
with subsequent propagation of the orbital flip.
In the latter case, the excitation process is as discussed above in the local crystal-field limit, i.e., it requires
the simultaneous excitation of a symmetry-breaking phonon to obtain a finite dipole moment for the local orbital flip.
The difference arises from the propagation of the orbital flip, which is similar to a spin flip which translates
into a spin wave or magnon in a system with long-range magnetic order. For such a propagating orbital flip we have
to take into account the dispersion. Due to momentum conservation, $\sigma(\omega)$ is sensitive to excitations with
$\Delta {\bf k}_{\rm tot}$\,=\,0, where $\Delta {\bf k}_{\rm tot}\,=\,{\bf k}_{\rm phonon}+{\bf k}_{\rm orbiton}$.
Since the symmetry-breaking phonon may have any momentum, $\sigma(\omega)$ reflects orbiton
contributions from the entire Brillouin zone. However, without more detailed theoretical predictions about the
line shape or width, this one-orbiton-plus-phonon peak cannot be distinguished from the broad vibronic Franck-Condon
peak expected in the local CF limit. Thus, we can attribute the absorption feature around 0.2\,eV to a single orbital
flip, but we cannot decide whether this flip is only local or propagates due to orbital exchange interactions.

In the other case mentioned above, the excitation process is based on the exchange of orbitals between adjacent sites.
Here, we primarily focus on the peak for $E$\,$\parallel$\,$c$ at 0.4\,eV.\@ Our interpretation of this
feature as a two-orbiton excitation naturally explains its energy, the polarization dependence,
and the pronounced temperature dependence, i.e., it resolves the three issues mentioned above.
For the exchange along the $c$ axis, we consider only the dominant processes and neglect
the rotation and tilt distortions of the octahedra. In this case, hopping preserves the type of orbital,
i.e., the only finite hopping processes are $xz(\alpha)\leftrightarrow xz(\beta)$ and
$yz(\alpha)\leftrightarrow yz(\beta)$, where $\alpha$ and $\beta$ denote neighboring V(1) and V(2) sites,
respectively (see Fig.\ \ref{fig:TwoOrbiton}). Note that hopping in $z$ direction is zero for the $xy$ orbital,
which is the lowest orbital on all sites and thus occupied by one electron.
Following the analysis of synchrotron x-ray data by Noguchi {\it et al.},\cite{Noguchi2000}
we consider \emph{G}-type orbital order for the intermediate phase (see Fig.~\ref{Fig:YVO3orbitalOrdering}),
i.e., the second electron per site occupies $xz$ on V(1) and $yz$ on V(2) in the ground state.
For this second electron per site we consider the fermionic Hamiltonian $H_F$
\begin{subequations}
\label{eq:fermions}
\begin{eqnarray}
\label{eq:hopping}
H_F &=& H_{F0} + t (c^\dagger_{\alpha1} c^{\phantom\dagger}_{\beta0}
+ c^\dagger_{\alpha0} c^{\phantom\dagger}_{\beta1} +{\rm h.c.}
)\\
\nonumber
H_{F0} &=& \sum_{\tau\in\{\alpha,\beta\}} \left(
(\varepsilon_\tau + \Delta_\tau) c^\dagger_{\tau1} c^{\phantom\dagger}_{\tau1}
+\varepsilon_\tau c^\dagger_{\tau0} c^{\phantom\dagger}_{\tau0}
\right.
\\
& & + \left.
U' c^\dagger_{\tau1} c^{\phantom\dagger}_{\tau1}
c^\dagger_{\tau0} c^{\phantom\dagger}_{\tau0} \right)
\end{eqnarray}
\end{subequations}
where $t$ denotes the hopping matrix element, $c^\dagger_{\tau i}$ ($c_{\tau i}$) creates (annihilates)
an electron in orbital $i$\,=\,0 or 1 on site $\tau\in\{\alpha,\beta\}$,
$U^\prime=U - 3J_H$ denotes the Coulomb repulsion for two electrons on the same site but in different
orbitals, and the energies $\varepsilon_\tau$ and $\Delta_\tau$ are illustrated in Fig.\ \ref{fig:TwoOrbiton}.
No spin appears since we consider parallel spins due to Hund's coupling to the electron in the $xy$ orbital.
In the orbital ground state, the lower levels (denoted $0$) are occupied by one electron, the upper levels
(denoted $1$) are empty.
The creation of an orbiton at site $\tau$ corresponds to the excitation of an electron from 0 to 1 at $\tau$,
which requires the energy $\Delta_\tau$. We introduce bosonic orbiton creation operators
$b_\tau^\dagger$ with
\begin{equation}
b_\tau^\dagger := c^\dagger_{\tau1} c^{\phantom\dagger}_{\tau0}.
\end{equation}
The annihilation operators are the hermitean conjugate ones. Note that
the orbitons are hardcore bosons since at maximum there can be only one
at each site.

In second order in $t$ the orbital ground state of $H_F$ (state (a) in Fig.\ \ref{fig:TwoOrbiton}) is linked
via the intermediate states (i) or (ii) to state (b) with {\em two} excited orbitons, one at site $\alpha$ and
one at site $\beta$. Note that states (c) and (d) corresponding to the excitation of a single orbiton cannot
be reached from the ground state in the considered symmetry.\cite{one}
In contrast, the two-orbiton excitation can account for the observed energy, polarization and temperature dependence.
For the energy of a two-orbiton excitation, one roughly expects twice the energy of a one-orbiton
excitation, neglecting orbiton-orbiton interactions and kinetic effects. As discussed above,
single orbital excitations are observed around 0.2\,eV, hence the energy of the feature at
0.4\,eV is well described by a two-orbiton interpretation.
Due to Hund's coupling with the electron in the low-lying $xy$ orbital, this two-orbiton excitation
requires parallel alignment of the spins on the considered bond, which in YVO$_3$ is only present
in the intermediate phase and only along the $c$ axis, in agreement with the observed polarization
and temperature dependence.
Moreover, the two-orbiton excitation requires hopping from, e.g., $xz$ on V(1) to $xz$ on V(2),
thus the spectral weight becomes zero if both orbitals are occupied in the ground state.
This is the case for the C-type OO observed below $T_S$, explaining the abrupt drop of the intensity at $T_S$.
For the intermediate phase, we have assumed pure G-type OO for simplicity, neglecting admixtures of C-type
(see discussion in the introduction).
Note that deviations from G-type OO as claimed recently on the basis of LDA+DMFT calculations\cite{deRay07}
will only affect the precise value of the spectral weight. In order to understand the temperature dependence
at higher temperatures, we now address the dipole-selection rule and calculate the spectral weight
of this two-orbiton excitation.

In analogy to the derivation of Heisenberg exchange, we derive an effective Hamiltonian $H_{\rm orb}$ in terms
of orbiton creation and annihilation operators for the mixture of states (a) and (b)
\begin{equation}
\label{eq:bosons}
H_{\rm orb} = J (b^\dagger_\alpha b^\dagger_\beta + {\rm h.c.}) + {\rm const}.
\end{equation}
In this effective description no virtual double occupancies appear.
To do so we assume that the on-site Coulomb interaction is the largest energy
\begin{subequations}
\label{eq:assume}
\begin{eqnarray}
U' & > & |\varepsilon_\alpha + \Delta_\alpha - \varepsilon_\beta|\\
U' & > & |\varepsilon_\beta + \Delta_\beta - \varepsilon_\alpha| \, .
\end{eqnarray}
\end{subequations}
This assumption is certainly met in YVO$_3$ where $U' =  U - 3J_H \approx 2$\,eV
and all other energies are of the order of fractions of an eV.\@
We find (see Appendix)
\begin{equation}
J = \frac{2t^2U'}{(U')^2-(\delta - ecE)^2} \, ,
\label{eq:J}
\end{equation}
where $e$ is the elementary charge, $c$ the distance between the two sites,
$E$ denotes an electric field applied parallel to the bond along the $c$ axis,
and
\begin{equation}
\delta = \varepsilon^0_\beta - \varepsilon^0_\alpha + \frac{\Delta_\beta-\Delta_\alpha}{2}
= (\varepsilon^0_\beta + \varepsilon^1_\beta - \varepsilon^0_\alpha - \varepsilon^1_\alpha)/2 \, .
\label{eq:delta}
\end{equation}
The two-orbiton excitation is dipole allowed if $\partial H_{\rm orb}/\partial E$\,$\neq$\,0.
We find
\begin{equation}
\partial J/\partial E \approx -\frac{4t^2}{(U')^3} \, \delta \, ec \, .
\label{eq:dJdE}
\end{equation}
In the denominator we neglected $\delta$ because it is much smaller than $U'$.

In the presence of a mirror plane between the two V sites, i.e., $\varepsilon^0_\beta = \varepsilon^0_\alpha$
and $\Delta_\beta = \Delta_\alpha$, the two-orbiton excitation does not carry a dipole moment and does
not contribute to $\sigma(\omega)$. However, such a
mirror plane is present only below $T_S$\,=\,77\,K and above $T_{OO}$\,=\,200\,K, but the symmetry is broken
in the intermediate phase with two distinct V sites. The situation is similar to the case of two-magnon absorption
discussed by Lorenzana and Sawatzky.\cite{Lorenzana1995} They demonstrated that two-magnon absorption becomes weakly
infrared active if the mirror symmetry on the bond is broken by the simultaneous excitation of a phonon.
In the present case, the symmetry is already broken without a phonon.
We conclude that the excitation of two orbitons is directly infrared active for $E$\,$\parallel$\,$c$
in the intermediate, monoclinic phase of YVO$_3$.

The absolute value of $\partial J/\partial E$ depends on the crystal-field levels via $\delta$.
From the first-principles study by Solovyev\cite{Solovyev06} and the LDA+DMFT calculation by
De Raychaudhury {\em et al.}\cite{deRay07} we obtain $\delta$\,=\,20 and 33.5\,meV, respectively.
However, each individual level $\varepsilon^i_\tau$ certainly is known only to within 50\,meV,
thus the error of $\delta$ is about 100\,meV.\@
For $U'= 2$\,eV, $\delta=20$ -- 100\,meV, and $t=100$ -- 150\,meV (Ref.\ \onlinecite{deRay07})
we obtain $\partial J / \partial E \, = \, 0.1$ -- $1.1 \cdot 10^{-3}ec$.
The spectral weight usually is expressed in terms of the plasma frequency $\omega_p$,
\begin{equation}
\int_{\omega_{\rm lo}}^{\omega_{\rm hi}} \sigma_1(\omega) d\omega = \frac{\pi \varepsilon_0}{2}\omega_p^2 \, ,
\label{eq:omegap}
\end{equation}
where the frequency range of interest is defined by $\omega_{\rm lo}$ and $\omega_{\rm hi}$,
and $\varepsilon_0$ denotes the dielectric constant of vacuum. The spectral weight of the two-orbiton excitation
at $\hbar \omega_{2o}$\,=\,0.4\,eV is given by
\begin{equation}
\omega_p^2 = \frac{2 \omega_{2o}}{\hbar \varepsilon_0 V} \, |\partial J /\partial E|^2 \, ,
 \label{eq:specweight}
\end{equation}
where $V$\,=\,56\,\AA$^3$ is the volume per site.
Finally we obtain $\hbar \omega_p \, = \, 0.6$ -- 6.8\,meV.\@
This has to be compared with the experimental result for the spectral weight, for which we choose
$\hbar \omega_{\rm lo}$\,=\,0.20\,eV and $\omega_{\rm hi}$\,=\,0.68\,eV.\@ In order to separate the two-orbiton
contribution from a background of, e.g., multi-phonon absorption, the integration for each temperature $T$ is
performed over $\sigma_1(T,\omega)-\sigma_1(4 {\rm K},\omega)$, since the two-orbiton excitation is not dipole
allowed at 4 K (see Fig.\ \ref{fig:weightc}). At 80\,K this yields $\hbar \omega_p^{\rm exp}= 16$ meV,
about a factor of 2 -- 26 larger than the calculated result.
We emphasize that the quantitative prediction of the spectral weight is a challenging task. Note that
the theoretical estimate of $\omega_p$ of the phonon-assisted two-magnon absorption in the cuprates
on the basis of a similar perturbation expansion was off by a factor of 4 -- 7.\cite{Lorenzana1995,thesisGrueninger}
Therefore we consider this result as a clear support for our interpretation.

The temperature dependence of $\omega_p$ is plotted in the inset of Fig.\ \ref{fig:weightc}.
Upon warming above $T_{OO}$\,=\,200\,K, the mirror plane is restored and the direct contribution
to $\sigma(\omega)$ is suppressed. However, the spectral weight above $T_{OO}$ is larger than at 10\,K.\@
This may either arise from a weak, phonon-assisted two-orbiton contribution or from thermal broadening
of the electronic gap and of the absorption band around 0.2\,eV (see Fig.\ \ref{fig:weightc}).
In the case of phonon-assisted two-{\it magnon} absorption proposed by Lorenzana and Sawatzky,\cite{Lorenzana1995}
$\sigma(\omega)$ in low-dimensional antiferromagnets is hardly affected at the magnetic ordering
temperature $T_N$, because the spin-spin correlation length remains large above $T_N$.
In YVO$_3$, measurements of the specific heat\cite{Blake2002} show that only of the order of 10\% of the expected entropy
are released in the phase transition at $T_{OO}$, which indicates strong fluctuations.
Therefore, a finite contribution of phonon-assisted two-orbiton excitations is possible above $T_{OO}$.
In contrast, the spectral weight abruptly vanishes upon cooling below $T_S$\,=\,77\,K, because the
mirror symmetry is restored and because both the orbital and the magnetic ordering patterns
change (see Fig.\ \ref{Fig:YVO3orbitalOrdering}).
The reduction of $\hbar \omega_p$ from 16 to 10\,meV observed between $T_S$ and $T_{OO}$ can tentatively
be attributed both to a reduction of ferromagnetic correlations between nearest neighbors above $T_N$ and
to rather small changes of the orbital occupation or of the crystal-field levels.
For $\delta \, = \, 20$ -- 100\,meV, a reduction of $\omega_p$ by a factor of 1.6 corresponds to
changes of the four individual CF levels of only about 4 -- 20\,meV.

The line shape in principle may serve as a key feature to test our interpretation. However, this requires to
take into account the orbital exchange interactions and the coupling to the lattice on the same footing.\cite{Schmidt2007}
On top of that, also orbiton-orbiton interactions have to be considered. Up to now, theoretical predictions
for the two-orbiton contribution to $\sigma(\omega)$ are not available. Due to momentum conservation,
$\sigma(\omega)$ is restricted to the observation of two-orbiton processes with
${\bf k}_{\rm tot}$\,=\,${\bf k}_1$\,+\,${\bf k}_2$\,$\approx$\,0. This means that the two orbitons have
opposite momenta ${\bf k}_1$\,=\,$-{\bf k}_2$ with arbitrary ${\bf k}_i$, thus the orbiton dispersion is
probed throughout the entire Brillouin zone. The total line width is a convolution of twice the
orbiton band width and the width arising from vibronic coupling to the lattice.

\begin{figure}[tb]
\includegraphics[clip,width=0.95\linewidth]{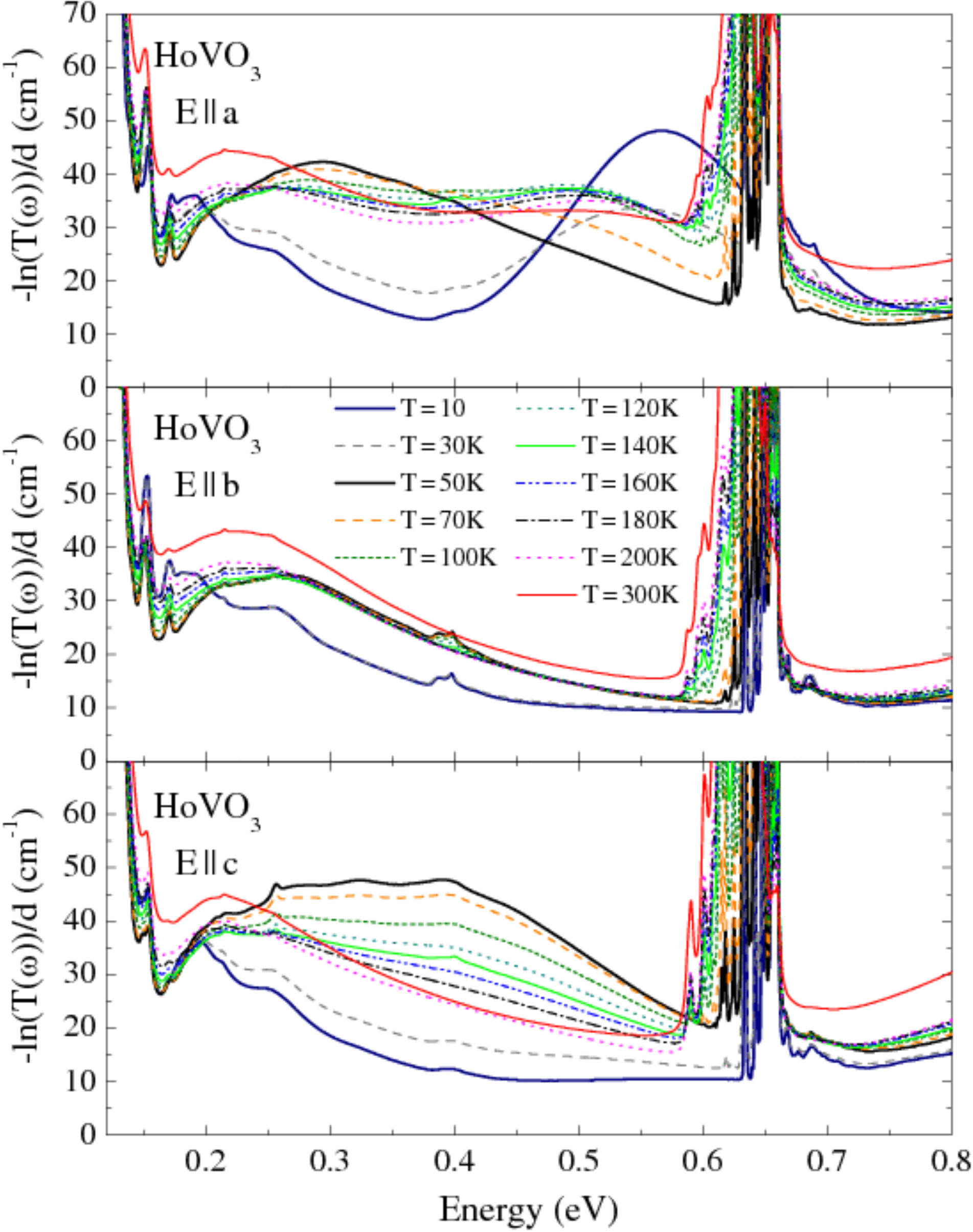}
\caption{(color online)
Temperature dependence of $-\ln({\rm T})/d$ of HoVO$_3$
for $E$\,$\parallel$\,$a$ (top panel), $E$\,$\parallel$\,$b$ (middle), and $E$\,$\parallel$\,$c$ (bottom).
The band of sharp lines at 0.6\,eV originates from crystal-field excitations within the Ho $4f$ shell.
} \label{fig:HoVO3}
\end{figure}

\begin{figure}[tb]
\includegraphics[clip,width=0.95\linewidth]{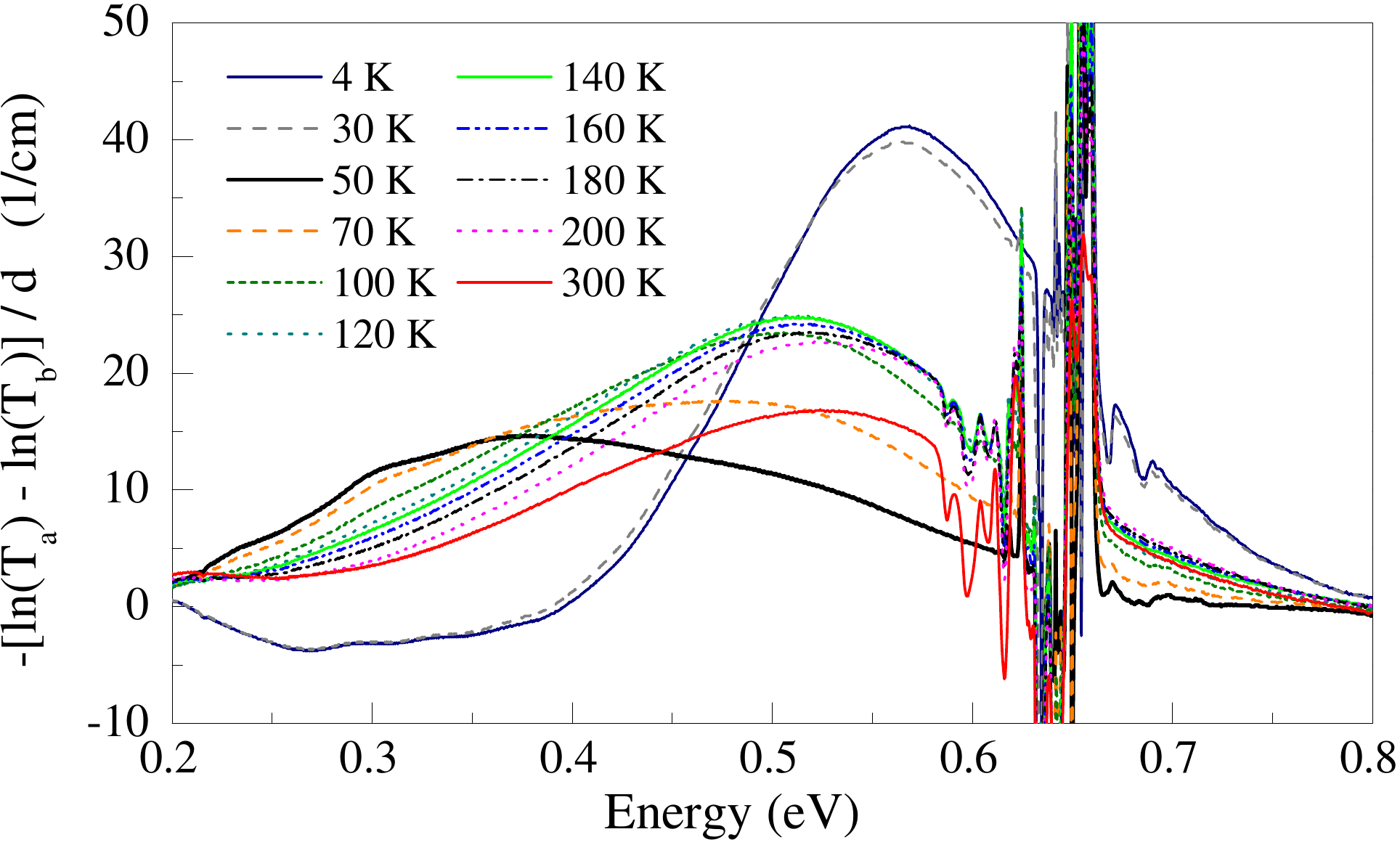}
\caption{(color online)
Anisotropy of the absorption spectra for $E$\,$\parallel$\,$a$ and $E$\,$\parallel$\,$b$ for HoVO$_3$. }
\label{fig:HoVO3aniso}
\end{figure}

\begin{figure}[tb]
\includegraphics[clip,width=0.6\linewidth]{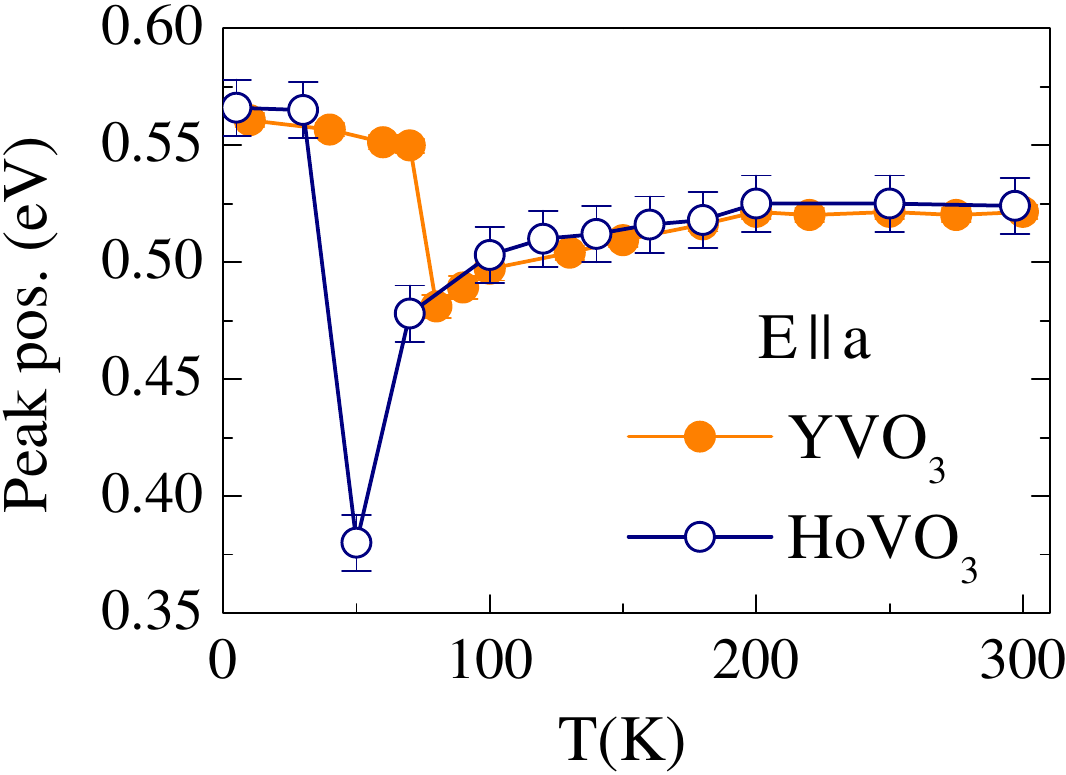}
\caption{(color online) Peak frequency observed for $E$\,$\parallel$\,$a$ in YVO$_3$ and HoVO$_3$.
} \label{fig:Posa}
\end{figure}

In order to test our interpretation of the feature in YVO$_3$ at 0.4\,eV for $E\,\parallel \,c$, we study HoVO$_3$,
which shows a very similar crystal structure and very similar magnetic and orbital ordering patterns.
The absorption coefficient $\alpha(\omega) \propto -\ln{({\rm T}(\omega))}/d$ of HoVO$_3$ shows qualitatively the
same mid-infrared features as observed in YVO$_3$, i.e., a peak at 0.4\,eV for $E$\,$\parallel$\,$c$ in the
intermediate phase, and a peak at about 0.55\,eV for $E$\,$\parallel$\,$a$ in the whole temperature range with the
same anisotropy between $\sigma_a(\omega)$ and $\sigma_b(\omega)$ (see Fig.\ \ref{fig:HoVO3}).
In HoVO$_3$, the phase transitions occur at $T_S$\,$\sim$\,40\,K and $T_{OO}$\,$\sim$\,188\,K.\cite{Blake2008}
Additionally, an absorption band consisting of several sharp lines is observed around 0.64\,eV, which can be attributed
to the $f$-$f$ transitions of the Ho$^{3+}$ ion ($^5I_8$\,$\rightarrow$\,$^5I_7$). The fine structure of this band
is due to transitions between different Stark components. Note that the polarization of the incident light mainly affects
the intensity of the observed lines.
The close similarity of the spectra of YVO$_3$ and HoVO$_3$ regarding the orbital excitations within the $3d$ shell
and in particular the sensitivity to the phase transitions clearly show that the considered features reflect intrinsic
properties of the vanadates and corroborate our interpretation.

What are the implications of our results for the claimed observations of orbitons by Raman
scattering\cite{Miyasaka2005,Miyasaka2006,Sugai2006} at energies of the order of 40 -- 80\,meV (see the introduction)?
The excitation energy of 0.4\,eV for $E\!\parallel \! c$ implies that $\Delta_\alpha + \Delta_\beta  \approx 0.4$\,eV.\@
This may be consistent with a comparably small one-orbiton energy if the CF splitting on V(1) and V(2) differ
substantially, e.g.\ $\Delta_\alpha \, = \,50$ -- 100\,meV and $\Delta_\beta \, = \,300$ -- 350\,meV.\@
Note that a rather large value of $\Delta_\beta - \Delta_\alpha$ is in agreement with the observed
value of the spectral weight (see above). However, the spectra for $E\! \parallel \! b$ indicate
$\Delta_\tau < 300$\,meV, implying $\Delta_\tau > 100$\,meV.

Finally, we turn to the feature at 0.55\,eV for $E$\,$\parallel$\,$a$.
Its rather high energy suggests that excitations from the low-lying $xy$ orbital are involved.
At all temperatures studied here, there is no significant contribution to $\sigma_b(\omega)$ around 0.5\,eV
(see Figs.\ \ref{fig:YVO3} and \ref{fig:HoVO3}).
This pronounced anisotropy puts severe constraints on the interpretation.
Assuming an ideal perovskite structure, one expects $\sigma_a(\omega)\,=\,\sigma_b(\omega)$.
In YVO$_3$ and HoVO$_3$, the V-O-V bonds are rotated within the $ab$ plane by about $45^\circ$ with respect
to the orthorhombic $a$ and $b$ axes,
with antiferro-orbital ordering in the entire temperature range (see Fig.\ \ref{Fig:YVO3orbitalOrdering}).
Therefore, one expects that an exchange process between {\em neighboring} V sites contributes roughly equally
to $\sigma_a(\omega)$ and $\sigma_b(\omega)$. In contrast, {\em next-nearest} V neighbors are displaced parallel to
the $a$ or $b$ axes. One may speculate that the structural distortions, i.e., rotations and tilts of the octahedra,
give rise to the observed anisotropy. Such a detailed theoretical analysis of the exchange between next-nearest
neighbors including structural details is beyond the scope of the present paper.
In order to highlight the anisotropy and to obtain a better view of the line shape, we plot the difference spectra
between $a$ and $b$ axis of HoVO$_3$ in Fig.\ \ref{fig:HoVO3aniso}. The temperature dependence of the peak frequency
is shown in Fig.\ \ref{fig:Posa}. Upon cooling down from 300\,K, the peak position, spectral weight and line shape
all change below about 100\,K.\@ This is more evident in HoVO$_3$, because the change is cut off at the first-order
transition at $T_S$, which is higher in YVO$_3$. Possibly, this may be related to the Ne\'{e}l temperature, $T_N$\,=\,116\,K
in YVO$_3$ and 114\,K in HoVO$_3$.\cite{Blake2008} This points towards the entanglement of spin and orbital degrees
of freedom, which is expected if exchange interactions are dominant.\cite{Oles06}

\section{Conclusion}

In conclusion, the orbital excitations in YVO$_3$ and HoVO$_3$ have been studied by optical spectroscopy.
We focused on an absorption band observed at 0.4\,eV for $E \! \parallel \! c$. We have shown that this feature
is located far below the Mott-Hubbard gap and that it can neither be interpreted in terms of phonons, magnons,
weakly bound (Mott-Hubbard) excitons, or polaronic carriers trapped at impurity sites.
Therefore, we identify this feature as an orbital excitation. However, based on the comparison with the
data of VOCl and with recent calculations,\cite{deRay07,Solovyev06} we have shown that this absorption peak
cannot be explained in a local crystal-field scenario, i.e., within single-site physics.
Alternatively, we propose that this peak reflects collective orbital excitations, i.e.,
orbital excitations that are based on the exchange coupling between neighboring V sites. We demonstrate
that the exchange of two orbitals between adjacent sites along the $c$ axis in the intermediate phase
directly contributes to $\sigma(\omega)$. The energy, polarization and temperature dependence as well as the
spectral weight of the absorption band at 0.4\,eV are in excellent agreement with the expectations for a
two-orbiton excitation.

Our results call for a number of further investigations. Our claim can be tested directly by the direct observation
of the dispersion with a momentum-resolving technique such as resonant inelastic x-ray scattering or electron
energy loss spectroscopy. Moreover, we call for theoretical studies of the orbital exchange that realistically
take into account the coupling to the lattice. In particular, a comparison of our data with predictions for
the line shape of two-orbiton absorption is expected to reveal important information on orbital-orbital interactions.
Finally, more detailed investigations of the exchange between next-nearest neighbors within the $ab$ plane
are necessary to clarify the nature of the absorption band observed at 0.55\,eV for $E\parallel a$.

\section{Appendix}

Here, we discuss the derivation of the effective Hamiltonian $H_{\rm orb}$ (see Eq.\ \ref{eq:bosons}).
Conceptually, this is even in second order less trivial than one might think at first glance.
This is so because the states without double occupancy (states (a) and (b) in Fig.\ \ref{fig:TwoOrbiton})
are not degenerate due to the differences in the energies $\varepsilon_\tau$ and $\Delta_\tau$.
This leads to the remarkable phenomenon that different second order calculations lead
to different results. This stems from the different ways
to perform the unitary transformation which eliminates the terms
which change the number of double occupancies. Similar observations
were made previously in the derivation of the electron-electron
attraction mediated by phonons \cite{lenz96}. We illustrate this
issue here by two calculations.

Both calculations require to split the hopping part of $H_F$
(second term in Eq.\ \eqref{eq:hopping}) in two parts
\begin{equation}
t (c^\dagger_{\alpha1} c^{\phantom\dagger}_{\beta0}
+ c^\dagger_{\alpha0} c^{\phantom\dagger}_{\beta1} +{\rm h.c.}
)
=
H_{F+}+ H_{F-}
\end{equation}
with $H_{F-}=(H_{F+})^\dagger$ and $H_{F+} = H_{F++} + H_{F-+}$.
The first plus (minus) sign indicates that a double occupancy is
created (annihilated). The second sign indicates whether an electron
is raised (+) or lowered ($-$), i.e., hops from $0$ to $1$ (+) of vice versa
($-$).
This implies $H_{F-+}=(H_{F+-})^\dagger$ and $H_{F--}=(H_{F++})^\dagger$.
In detail, we have
\begin{subequations}
\begin{eqnarray}
H_{F++} &=& t(c^\dagger_{\alpha1} c^{\phantom\dagger}_{\beta0}
\hat n_{\alpha0} +c^\dagger_{\beta1} c^{\phantom\dagger}_{\alpha0}
\hat n_{\beta0})\\
H_{F+-} &=& t(c^\dagger_{\alpha0} c^{\phantom\dagger}_{\beta1}
\hat n_{\alpha1} +c^\dagger_{\beta0} c^{\phantom\dagger}_{\alpha1}
\hat n_{\beta1}).
\end{eqnarray}
\end{subequations}

\subsection{Standard Unitary Transformation}
The standard approach is to determine an antihermitean operator $\eta=\eta_+ - \eta_-$ such that
\begin{equation}
\label{eq:onestep}
H_{\rm orb} = \exp(\eta)H_F\exp(-\eta)
\end{equation}
holds. To eliminate the hopping in leading order we require $[\eta,H_{F0}]=-H_{F+}-H_{F-}$ which leads to
\begin{subequations}
\begin{eqnarray}
\eta_{++} &=& \frac{t
c^\dagger_{\alpha1} c^{\phantom\dagger}_{\beta0} \hat n_{\alpha0}}
{U'+\Delta_\alpha-\Delta\varepsilon}
 +
\frac{t c^\dagger_{\beta1} c^{\phantom\dagger}_{\alpha0}
\hat n_{\beta0}}{U'+\Delta_\beta+\Delta\varepsilon}
\\
\eta_{+-} &=& \frac{tc^\dagger_{\alpha0} c^{\phantom\dagger}_{\beta1}
\hat n_{\alpha1}}{U'-\Delta_\beta-\Delta\varepsilon}
 +
\frac{tc^\dagger_{\beta0} c^{\phantom\dagger}_{\alpha1}
\hat n_{\beta1}}{U'-\Delta_\alpha+\Delta\varepsilon}
,
\end{eqnarray}
\end{subequations}
where we used $\Delta \varepsilon=\varepsilon_\beta-\varepsilon_\alpha$, and
$\eta_+=\eta_{++} + \eta_{+-}$ as for the parts of the Hamiltonian, with $\eta_- = \eta_{--} + \eta_{-+}$,
$\eta_{--}=(\eta_{++})^\dagger$, and $\eta_{-+}=(\eta_{+-})^\dagger$.
In second order in $t/{\cal O}(U')$ we obtain
$H_{\rm orb} = \frac{1}{2}[\eta,H_{F+}+H_{F-}]$. Using the shorthand
$H_{\rm orb +}=Jb^\dagger_\alpha b^\dagger_\beta$
for the creation of orbitons we have to compute
\begin{equation}
\label{eq:commutate-standard}
H_{\rm orb +}= (1/2)\left( [\eta_{++},H_{F-+}]+ [H_{F++},\eta_{-+}] \right).
\end{equation}
Explicit commutation leads to the standard result
\begin{equation}
\label{eq:standard}
J_{\rm stan}=\frac{t^2 U'}
{(U')^2-(\Delta_\alpha-\Delta\varepsilon)^2}+
\frac{t^2 U'}
{(U')^2-(\Delta_\beta+\Delta\varepsilon)^2}.
\end{equation}
For $\Delta_\tau=0=\Delta\varepsilon$ this is identical to the result known
from the derivation of the Heisenberg spin exchange as in
$(J_{\rm Heisen}/2)(S_\alpha^+S_\beta^- +S_\alpha^-S_\beta^+)$
which implies $2J=J_{\rm Heisen}=4t^2/U'$.
Note that Eq.\ \eqref{eq:standard} for the exchange $J$ becomes singular as soon as
$U'\to|\Delta_\tau\pm\Delta\varepsilon|$. We will see that a smoother unitary transformation
provides a less singular result.

\subsection{Continuous Unitary Transformation (CUT)}

It might surprise that the result \eqref{eq:standard}
is not unique. But we emphasize that only the matrix elements
on-shell, i.e., without energy change, are defined independently from
the chosen basis. All other matrix elements do depend on the
chosen basis. Generally, a continuous change of basis
is smoother and less singular than the one-step transformation,
see also Refs.\ \onlinecite{wegne94,lenz96}.

The continuous change of the Hamiltonian is parameterized
by $\ell\in[0,\infty)$ and $H_F(\ell)$ is given by the differential
equation
\begin{equation}
\label{eq:cut}
\partial_\ell H_F(\ell) = [\eta(\ell),H_F(\ell)].
\end{equation}
It is understood that $H_F(\ell=0)$ is given by the Hamiltonian $H_F$ in \eqref{eq:fermions} while
$H_F(\ell=\infty)$ is given by $H_{\rm orb}$ in \eqref{eq:bosons}. The transformation \eqref{eq:cut}
shall eliminate the terms in $H_F$ which change the number of double occupancies,
i.e., the kinetic part $H_{F+}+H_{F-}$. Hence we parameterize
\begin{subequations}
\begin{eqnarray}
H_{F++}(\ell) &=&  A_1(\ell) c^\dagger_{\alpha1} c^{\phantom\dagger}_{\beta0}
\hat n_{\alpha0} + B_1(\ell) c^\dagger_{\beta1} c^{\phantom\dagger}_{\alpha0}
\hat n_{\beta0}\\
H_{F+-}(\ell) &=&  A_0(\ell) c^\dagger_{\alpha0} c^{\phantom\dagger}_{\beta1}
\hat n_{\alpha1} +
B_0(\ell) c^\dagger_{\beta0} c^{\phantom\dagger}_{\alpha1}
\hat n_{\beta1} , \qquad
\end{eqnarray}
\end{subequations}
while $H_{F0}$ remains constant in linear order in $t$; the operators
$H_{F--}(\ell)=(H_{F++}(\ell))^\dagger$ and
$H_{F-+}(\ell)=(H_{F+-}(\ell))^\dagger$ follow by hermitean conjugation.

The crucial choice is the one for the infinitesimal generator $\eta(\ell)$.
Our aim is to eliminate processes which create or annihilate excitations
of the order of $U'$. Such an elimination can most easily be done by
the Mielke-Knetter-Uhrig generator $\eta_{\rm MKU}(\ell)$
\cite{mielk98,uhrig98c,knett00a} which consists of the terms in the
Hamiltonian increasing the number of excitations and of the negative
terms in the  Hamiltonian decreasing the number of excitations
\begin{equation}
\eta_{\rm MKU}(\ell) := H_{F+}(\ell)- H_{F-}(\ell),
\end{equation}
for a general discussion see also Ref.\ \onlinecite{knett03a}. With this choice
one obtains in linear order in $t$
\begin{equation}
\partial_\ell H_{F+}(\ell) = -[H_{F+}(\ell),H_{F0}]
\end{equation}
which implies the differential equations
\begin{subequations}
\label{eq:diffeq}
\begin{eqnarray}
\partial_\ell A_1 &=& -(U'+\Delta_\alpha-\Delta\varepsilon)A_1\\
\partial_\ell A_0 &=& -(U'-\Delta_\beta-\Delta\varepsilon)A_0\\
\partial_\ell B_1 &=& -(U'+\Delta_\beta+\Delta\varepsilon)B_1\\
\partial_\ell B_0 &=& -(U'-\Delta_\alpha+\Delta\varepsilon)B_0.
\end{eqnarray}
\end{subequations}
The solutions consist in decreasing exponential functions starting
at $t$ for $\ell=0$ because we assume all the energy differences in the
parentheses in \eqref{eq:diffeq} to be positive, i.e., $U'$ dominates
the other energies, see Eq.\ \ref{eq:assume}.

The orbital exchange is obtained by equating the second order terms
in $t$ in \eqref{eq:cut} which implies
\begin{equation}
\label{eq:commutate-cut}
\partial_\ell J(\ell)  b^\dagger_\alpha b^\dagger_\beta =
2[H_{F++}(\ell),H_{F-+}(\ell)].
\end{equation}
Since the right hand side is given by the solutions of \eqref{eq:diffeq}
an integration suffices to provide $J_{\rm CUT}=J(\ell=\infty)$
\begin{subequations}
\label{eq:result}
\begin{eqnarray}
J_{\rm CUT} &=& 2 \int_0^\infty (A_1(\ell)A_0(\ell)+B_1(\ell)B_0(\ell))d\ell\\
 &=& \frac{t^2}{U'-\delta + ecE} + \frac{t^2}{U'+\delta -ecE} \\
&=& \frac{2t^2U'}{(U')^2-(\delta - ecE)^2} \, ,
\label{eq:cut-result}
\end{eqnarray}
\end{subequations}
where we use the shorthand $\delta$ for the crystal-field levels (see Eq.\ \ref{eq:delta}),
and $E$ denotes the applied electric field.
This is the result used in the main part of the article, see Eq.\ \ref{eq:J}.

Note that we retrieve the well-known result for the Heisenberg  exchange of
$2J=J_{\rm Heisen}=4t^2/U'$ for $E=0$ and $\delta=0$, i.e., equivalent V sites.

Even more interesting is that $J_{\rm CUT} \neq J_{\rm stan}$. In particular,
the individual excitation energies $\Delta_\tau$ do not occur in $J_{\rm CUT}$
in \eqref{eq:cut-result} but only their difference (see Eq.\ \ref{eq:delta}).
Hence a regime exists with $\delta=0$ and $\Delta_\tau \to U'$ where $J_{\rm stan}$ diverges
while $J_{\rm CUT}$ remains unaffected. So the CUT result is less singular. Moreover, the
expression for $J_{\rm CUT}$ is simpler than the one for $J_{\rm stan}$.

Tracing back from where the difference between the standard and the CUT results originates
we have to compare Eqs.\ (\ref{eq:commutate-standard}) and (\ref{eq:commutate-cut}). In the standard calculation
\eqref{eq:commutate-standard} there is a striking asymmetry between
the two operators which are commutated. Only one of them ($\eta$)
carries information on the excitation energies. In the CUT calculation
\eqref{eq:commutate-cut} both commutated operators carry the dependence
on the excitation energies in the same way by their
dependence on $\ell$. This implies also that
a single commutation suffices because two commutators are equal
while there are two different ones in \eqref{eq:commutate-standard}.

For all the above reasons we favor the CUT derivation.
We stress, however, that in the regime relevant for YVO$_3$
the difference between  $J_{\rm stan}$ and $J_{\rm CUT}$
is quantitatively of minor importance.

\section*{Acknowledgments}
It is a pleasure to acknowledge fruitful discussions with D.~I.~Khomskii, M.~Mostovoy, and P.~van Loosdrecht.
This project is supported by the DFG via SFB 608.


\begin{thebibliography}{50}
\bibitem{one}
From the ground state (a), the one-orbiton states (c) and (d) can only be reached if the off-diagonal hopping
$xz(\alpha)\leftrightarrow yz(\beta)$ is finite, i.e., it requires deviations from the ideal perovskite structure.
Therefore, the matrix element is significantly smaller than for the two-orbiton excitation.
\end{thebibliography}
\end{document}